\documentclass[aps,prb,superscriptaddress,showpacs,twocolumn]{revtex4}
\usepackage{graphicx}
\begin{document}
\title{Vortex Motion Noise in Micrometre-Sized Thin Films of the Amorphous Nb$_{0.7}$Ge$_{0.3}$
Weak-Pinning Superconductor}
\author{D. Babi\'{c}}
\altaffiliation[Corresponding author. Present address:  ] {Department of Physics, Faculty of Science,
University of Zagreb, Croatia}
\email{dbabic@phy.hr}
\affiliation{Institute of Physics, University of Basel, Klingelbergstrasse 82,
CH-4056 Basel, Switzerland}
\author{T. Nussbaumer}
\affiliation{Institute of Physics, University of Basel, Klingelbergstrasse 82,
CH-4056 Basel, Switzerland}
\author{C. Strunk}
\altaffiliation[Present address:  ] {Institute for Experimental and Applied Physics,
University of Regensburg, Germany}
\affiliation{Institute of Physics, University of Basel, Klingelbergstrasse 82,
CH-4056 Basel, Switzerland}
\author{C. Sch\"{o}nenberger}
\affiliation{Institute of Physics, University of Basel, Klingelbergstrasse 82,
CH-4056 Basel, Switzerland}
\author{C. S\"{u}rgers}
\affiliation{Physikalisches Institut, Universit\"{a}t Karlsruhe, D-76128 Karlsruhe, Germany}
%
%
%
%
\begin{abstract}
We report high-resolution measurements of voltage ($V$) noise in the mixed state
of micrometre-sized thin films of amorphous Nb$_{0.7}$Ge$_{0.3}$, which is
a good representative of weak-pinning superconductors.
There is a remarkable difference between the noise below and
above the irreversibility field $B_{irr}$. Below $B_{irr}$,
in the presence of measurable pinning, the noise at small applied currents
resembles shot noise, and
in the regime of flux flow
at larger currents decreases with increasing voltage
due to a progressive ordering of the vortex motion.
At magnetic fields $B$ between $B_{irr}$ and the upper critical
field $B_{c2}$ flux flow is present already at vanishingly small currents.
In this regime the noise scales with $(1-B/B_{c2})^2 V^2$ and has a
frequency ($f$) spectrum of $1/f$ type.
We interpret this noise in terms of the properties
of strongly driven depinned vortex systems
at high vortex density.
\end{abstract}
%
%
%
\maketitle

\section{\label{intro} Introduction}

When set in motion by a current $I$, vortices in superconductors generate a voltage $V$.
The resulting $V(I)$ curve may be either non-linear, implying depinning phenomena, or linear,
indicating flux flow (FF). Such $V(I)$ characteristics
do not provide complete information on the
nature of vortex motion,
especially if the pinning is weak. This is a point where information
available from the voltage noise becomes a powerful indicator of the underlying physics.
The finding  \cite{gurp} that vortices moving as bundles
composed of $N$ magnetic-flux quanta $\phi_0$
may produce shot noise attracted considerable attention, and resulted in
extensive subsequent work which was eventually extended beyond a simple
shot-noise approach.\cite{clem}
Samples used
in these studies were mainly polycrystalline conventional superconductors
with appreciable pinning and non-linear $V(I)$ characteristics up to very close to
$B_{c2}$. Noise experiments have also been carried out on
high-$T_c$ superconductors,\cite{woltgens,danna,kim} which are 
in a "liquid" state of negligible pinning over a
large portion of  the magnetic field vs. temperature plane, displaying linear $V(I)$
curves. However,  the intricate anisotropic character of vortex matter in these
compounds is a serious obstacle to the understanding of the mechanisms
that contribute to voltage noise related to motion of
vortices against a weak pinning potential.

Thus, a number of phenomena in the
weak-pinning regime have remained largely unexplored from the
point of view of vortex motion noise. The same holds
for the noise properties in the depinned state, i.e. for $B >
B_{irr}$. For instance, the interplay of bulk pinning and surface
barriers,\cite{bl,geom} which are both obstacles for vortex
motion (and can be of similar strengths when pinning is weak),
has been studied mostly by analysing the $V(I)$ curves and the
magnetoresistance $R(B,T)$.\cite{richard} Similarly, dynamic
ordering of vortex motion has also been explored by measuring
the average transport properties.\cite{aarts} Noise measurements
can reveal effects which are beyond reach of measurements of the average
voltage.  For example, if pinning is absent the 
$V(I)$ is linear, but one could ask does this mean that vortices really move
completely "silently" or are
there some dynamic effects which introduce fluctuations in their
velocity? Moreover, it is known that shot noise probes
the properties of "granular magnetic-flux charge", $N \phi_0$,  
but the details of this process are still subject to 
discussion - especially if $N$ is small
(characteristic of weak pinning).

In this paper we present high-resolution noise measurements
which address the above topics.
We have chosen a system particularly suitable for such research, namely
Nb$_{0.7}$Ge$_{0.3}$ amorphous thin films of thickness $d$ comparable to the
coherence length $\xi$. These films are conventional, isotropic, weak-coupling $s$-wave
BCS superconductors in the dirty limit, and
for $\xi \sim d$ they exhibit an extended ($B$,$T$) range of easily-movable vortices.\cite{theun}
In contrast to the complicated situation in high-$T_c$ compounds, here vortices
can be considered as undeformed "cylinders" of a volume $\xi^2 d$ and the Ginzburg-Landau (GL)
parameters can be found straightforwardly.  We also note that our shaping the
samples in the form of {\it narrow wires} turned out to be crucial for observing the
overall properties of the noise, i.e. for both $B < B_{irr}$ and $B > B_{irr}$.

In the regime where
$V(I)$ and $R(B,T)$ still indicate the presence of pinning
we find a noise similar to that
in Ref.\onlinecite{habjoin}, i.e. which for small applied currents resembles
shot noise, being linear in $V$
and frequency independent at low frequencies,
and decreases for more ordered vortex motion at high currents.
Closer to $B_{c2}$, over a rather extended range, we find no evidence for pinning in
neither $V(I)$ or $R(B,T)$. The lower boundary of this region is therefore
taken as the irreversibility field $B_{irr}$.
The noise for $B_{irr} < B < B_{c2}$ is qualitatively different from that in the
presence of pinning. It
exhibits a $1/f$ frequency spectrum
and is quadratic in $V$. Moreover, it scales with $(1-B/B_{c2})^2 V^2$.
The monotonic increase with increasing $V$, and in particular the scaling
which involves no pinning dependent parameters, motivates us to
propose that the noise of this kind is a peculiar
property of strongly driven vortices at high vortex density.

\section{\label{exp} Experiment}

Our samples (20 nm thick) were produced by magnetron sputtering
of Nb and Ge on to oxidised silicon wafers through masks prepared
by electron-beam lithography, using a double-layer resist
(PMMA/PMMA-MA). The measurements  were carried out in a $^4$He
cryostat, above the $\lambda$-point of liquid helium. Voltage
noise, $V(I)$ and $R(B,T)$ were measured extensively on a $W = 5$
$\mu$m wide and $L=50$ $\mu$m long wire connected to two wide
contact pads (sample S5). In order to investigate size effects in
the noise we performed a less comprehensive set of  measurements
on a $W=1$ $\mu$m  and $L=10$ $\mu$m sample (sample S1). By
analysing the low-current (10 nA; 10 Acm$^{-2}$)
$R(B,T)$ measurements within the framework of a model appropriate
for dirty weak-coupling superconductors \cite{kes} we
characterised sample S5 in detail. The transition temperature
$T_c = 2.91$ K  is determined as the midpoint 
of the 10 \% - 90 \% (0.1 K) zero-field transition curve. The transition curve is
smooth and free of "kinks" that would indicate the presence of
inhomogeneities, and we ascribe the rather wide transition (in
units of $T/T_c$) to a pronounced two-dimensional character of
the sample. A similar conclusion was drawn in
Ref.\onlinecite{dbjrc} for an YBa$_2$Cu$_3$O$_{7- \delta}$ single
crystal investigated systematically with respect to different
$\delta$-values and consequently different anisotropies. 
Very weak temperature dependence of the normal state resistivity 
$\rho_N$ above $T_c$ permits the estimation of
$\rho_N (T=0) = (2.3 \pm 0.2)$ $\mu \Omega$m.
Using this value and 
$ - (dB_{c2} / dT)_{T=T_c} \approx 2.05$ TK$^{-1}$, 
determined from the $R( B={\rm const.},
T)$ measurements (not shown), we calculate\cite{kes} the GL
parameters: $\xi (0) = 7.4$ nm, $\kappa = 77$ and $\lambda (0) =
1.63 \kappa \xi(0) = 930$ nm. The parameters of sample S5 are in
good agreement with published work.\cite{theun} Sample S1 had a
slightly lower $T_c$ (2.55 K) and larger $\rho_N$, but otherwise
showed fairly the same properties as sample S5. The method of
noise measurements is described in detail in
Ref.\onlinecite{henny}. In short, the signal from a sample is
processed through two low-noise amplifiers the outputs of which
are cross-correlated in a spectrum analyser. The noise setup is
calibrated against the equilibrium Nyquist noise $4 k_B T R_N$ in
the normal state ($R_N$ is the normal state resistance). By this
approach we have obtained a resolution of $\alt 10^{-20}$ V$^2$s,
necessary for measurements of small noise signals appearing in
the case of weak pinning. For both samples the frequency window
for the noise measurements was 106.5 - 114 kHz, except for the
measurements of the frequency dependence of  the noise power
spectrum $S_V$, performed at several frequencies between 20 kHz
and 250 kHz.

All the noise measurements were carried out at fixed
temperatures, $T=2.4$ K ($T/T_c = 0.82$) for sample S5 and
$T=2.25$ K ($T/T_c = 0.88$) for sample S1. 
Since sample S1 had lower $T_c$, we 
had to choose a larger value of $T/T_c$ 
in order to avoid temperature instabilities that appear in
the vicinity of the $\lambda$-point.

\section{\label{average} Magnetoresistance and current-voltage characteristics}

First we analyse the $R(B,T)$ and $V(I)$ results. In the lower
inset to Fig.1 we show $R(B , 2.4 \; {\rm K})$ for sample S5.
Above $\sim 0.65$ T we found good agreement with the FF theory of
Larkin and Ovchinnikov  (LO).\cite{lo} The LO FF conductivity is
given by 
\begin{equation}
\sigma_{FF} = \frac{1}{\rho_N}
\left[ 1+\frac{1}{(1-T/T_c)^{1/2}} 
\left( \frac{B_{c2}}{ B} \right) 
g \left(  B/B_{c2} \right)  \right]   ,
\label{LOFF}
\end{equation} 
where (for $z > 0.315$) $g(z) = (1 - z)^{3/2} [0.43+ 0.69 (1-z)]$. The solid
line, representing the LO FF resistance $R_{FF} = R_N /
\sigma_{FF} \rho_N$, is drawn by taking $T_c = 2.91$ K, $B_{c2}=
1.18$ T and $R [B_{c2}(2.4 \; {\rm K})] = R_N = 1375$ $\Omega$
($\rho_N = 2.75$ $\mu \Omega m$). The mentioned uncertainty in
$\rho_N$ implies a certain range of the $B_{c2}$ values that do
not deteriorate the fit. This range is $\sim 1.14 - 1.22$ T 
and agrees fairly well with $B_{c2} \sim 1.09 - 1.12$ T
obtained by the extrapolation method of Ref.\onlinecite{aarts}.
Henceforth we use $B_{c2} = 1.18$ T. Taking different values of
$T_c$ (within the transition width) has little effect on the
quality of the fit. We conclude that for the fields above $\sim
0.65$ T the vortices flow freely even at very small applied
currents, and thus $B_{irr} (2.4 \; {\rm K}) \sim 0.65$ T, which
is, as we show below, in agreement with the $V(I)$ results.

%
\begin{figure}
\includegraphics[width=75mm]{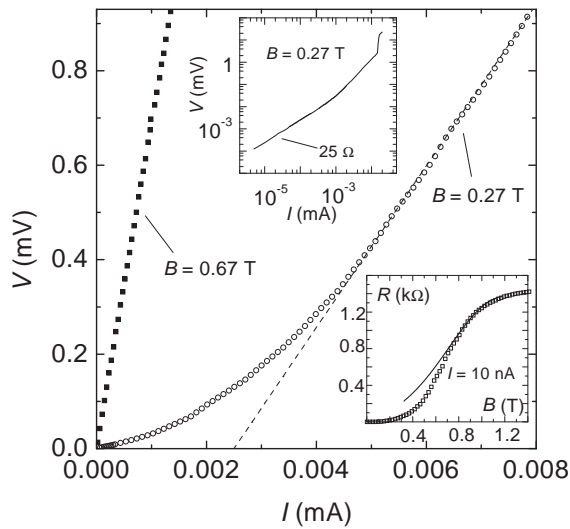}
\caption{$V(I)$ at 0.27 T (open circles) and 0.67 T (full
squares) for sample S5 at 2.4 K. At 0.27 T, for  large currents
there is a $V \propto (I - I_c)$ dependence  
(indicated by the dashed line).
At 0.67 T and higher fields the $V(I)$ are linear
starting from $I \rightarrow 0$ and over the whole range of our
noise measurements. Upper inset: Log-log plot of $V(I)$ at 0.27 T,
showing, at small currents, ohmic behaviour (with $R=25$ $\Omega$)
over two decades in $I$, and a jump to a value close to $V_N
=R_NI$ at high currents. Lower inset: $R(B, 2.4 \; {\rm K})$. The
solid line is the LO FF resistance drawn using $T_c = 2.91$ K,
$B_{c2} = 1.18$ T and $R_N = 1375$ $\Omega$.}
\end{figure}

For magnetic fields below $\sim 0.65$ T
the LO theory does not
explain the magnetoresistance data, and $R$ is smaller than $R_{FF}$. This indicates that the vortices
are slowed down by experiencing a pinning potential. However, $R$ is finite
even at magnetic fields as low as $\sim 0.05 B_{c2}$, which implies a very
weak pinning. In the upper inset to Fig.1 we show a log-log plot of a typical $V(I)$ in this region,
for 0.27 T.
Over two decades in $I$ the $V(I)$ is ohmic ($R = 25$ $\Omega$) before it turns
upwards. This suggests a hopping vortex motion (HVM), most probably
thermally activated.
In the  model of thermally activated HVM, vortex velocity is given by
$v_\phi = l (\nu_+ - \nu_-)$, where $l$ is the hop length and
$\nu_{\pm} \propto \exp[-(U \mp  U_F)/k_B T]$ the
hopping rates over a potential $U$ in the direction (+) and opposite (-) to the driving force
${\bf F} = - \nabla  U_F$. Since $F \propto IB$ and $V=BLv_\phi$,
for $I \rightarrow 0$ the $V(I)$ is linear.
From our measurements of  $R (B= {\rm const.}, T)$
we can estimate the values of $U/k_B$, which
are remarkably small.
At 2.4 K, $U/k_B$ is lower than 10 K and is a decreasing function of $B$.
At higher currents the $V(I)$  gradually changes to a $V (I) \propto(I - I_c)$
dependence, as we show for 0.27 T in Fig.1 by open circles.
This suggests a force-induced transition
to flux flow, i.e. an ordering of the vortex motion with increasing driving force.
This assumption will be supported further by the noise results
presented in Section \ref{lownoise}.
Finally, at even higher currents $V$ jumps to a value
of the order of $V_N = R_N I$ (Fig.1, upper inset) due to the appearance
of non-linear FF described in the LO theory \cite{lo} and observed
experimentally for similar films.\cite{aarts,lefloch}
Above 0.65 T, where FF takes place even at vanishingly small currents,
the $V(I)$ is simple:  linear starting from $I \rightarrow 0$
and all the way up to the appearance of non-linear effects in FF, as shown
for 0.67 T in Fig.1 by full squares.

\section{\label{noise} Noise results}

In the rest of the paper we present and discuss the results of our noise measurements,
which if not specified otherwise refer to sample
S5. We introduce $\Sigma_V$ to denote the excess noise, which is the
difference between the total measured noise $S_V$   and the thermal
(Nyquist) noise $4 k_B T (dV/dI)$.
The currents used in the noise measurements were always kept below those corresponding to
the appearance of the high-current non-linearities mentioned in Section \ref{average}, since we are
interested in situations where the average transport properties are still unaffected by the
high-current dynamical
processes described in the LO theory. \cite{lo}
In Section \ref{lownoise} we analyse the noise in the regime of non-linear $V(I)$
curves, i.e. for $B < B_{irr}$, and in Section \ref{highnoise} we turn to the noise for
$B_{irr}  < B <  B_{c2}$, where the $V(I)$ is linear and $R(B, 2.4 \; {\rm K})$ agrees well
with the LO FF theory.

\subsection{\label{lownoise} Noise in the regime of non-linear $V(I)$}

In Fig.2 we show a typical $\Sigma_V (V)$ curve in the regime of
non-linear $V(I)$, i.e. for 0.27 T (corresponding to the $V(I)$
curve in Fig.1). The maximum background Nyquist noise is $\sim
2.5 \times 10^{-20}$ V$^2$s. $\Sigma_V (V)$ first increases
linearly up to $V\simeq 0.2$ mV which is close to the upper limit
of HVM in $V(I)$. At higher voltages, where $V(I)$ becomes
proportional to $(I - I_c)$, $\Sigma_V$ gradually decreases with
increasing $V$. From this decrease of $\Sigma_V (V)$ we infer
that the vortex motion becomes more and more ordered when the
driving force progressively dominates over the pinning potential.
At large driving force the pinning potential causes not only a finite
offset $I_c$ in $V(I)$ but also random fluctuations of the vortex
velocity, which is most probably the origin of the small residual
noise above $V \sim 0.5$ mV. This residual noise is expected to
vanish together with $I_c$  at $B_{irr}$, which is indeed observed
in our experiment.  It is worthwile to note that the onset
of collective vortex motion has stronger effect on
$\Sigma_V$ than on $V(I)$. In $\Sigma_V$
the depinning threshold $I_c$ is indicated by a pronounced
maximum above which an ordering of the vortex motion occurs. On the
other hand, $V(I)$ shows no sharp feature at $I_c$, implying that
$I_c$ has to be determined by extrapolation of the linear
part of $V(I)$ down to $V=0$. Since the linear regime extends
only over a small current range between the HVM regime and the
high-current non-linearities, the determination of $I_c$ is
more ambiguous than in $\Sigma_V$. The non-monotonic character of
$\Sigma_V (V)$ supports our interpretation more strongly, and also
supplements research on dynamic vortex ordering
studied\cite{aarts} by analysing the average transport
properties.
%
%
%
\begin{figure}
\includegraphics[width=75mm]{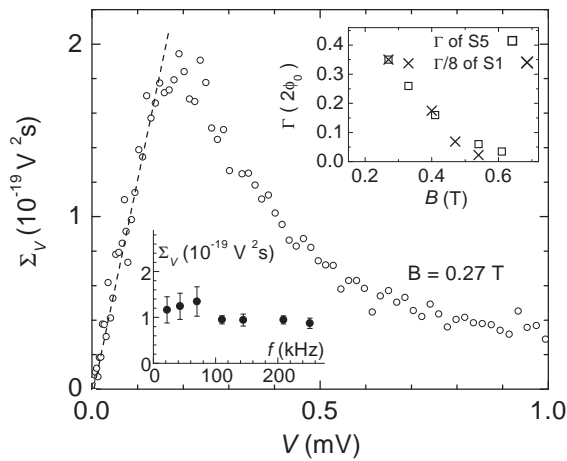}
\caption{Vortex motion noise $\Sigma_V(V)$ at 0.27 T and $T$= 2.4 K, corresponding to the $V(I)$
curve in Fig.1.
The dashed line indicates the linear $\Sigma_V (V)$ dependence.
Lower inset: $\Sigma_V (f)$ measured at 0.33 T and 1 $\mu$A (65 $\mu$V), in the linear part
of $\Sigma_V(V)$. Upper inset: Magnetic-field dependences of
the slopes $\Gamma$ (expressed in units of $2 \phi_0$) of the linear $\Sigma_V (V)$ curves.
The values for sample S1 divided by eight (crosses) agree well with those
for sample S5 (squares), which is in fair agreement with the assumption that
$\Gamma$ is inversely proportional to sample width.}
\end{figure}

Similarly to shot noise (in current) of electrons, which is a
linear function of $I$, shot noise (in voltage) of vortices is a
linear function of $V$. \cite{gurp} To check whether the linear
increase of $\Sigma_V (V)$ in the low-voltage regime can be
interpreted as shot noise we investigated the frequency and
sample-width dependences of $\Sigma_V$. The
studies\cite{gurp,habjoin,knoedler} of vortex-motion shot noise
offer different models for the slopes $\Gamma$ of linear
$\Sigma_V (V)$ plots, as we discuss later, but they agree in
predicting a frequency-independent $\Sigma_V(f)$ up to a
frequency $f_c \sim v_{\phi} /W = V/B L W$. {\it Because the wire
width W is small}, in the present case the (calculated) $f_c$ is
large, more than 500 kHz for all the measured points, except for
a few ones very close to $V = 0$. We measured $\Sigma_V(f)$ at a
characteristic point ($V = 65$ $\mu$V)
of a linear $\Sigma_V (V)$ curve, and found
that $\Sigma_V (f)$ is essentially flat between 20 kHz and 250
kHz, as shown in the lower inset to Fig.2. This result meets the
above-mentioned expectation for shot noise.

In the first work\cite{gurp} on vortex shot noise the factor
$\Gamma$ was related to the "charge" of a vortex bundle, i.e.
$\Gamma = 2 N \phi_0$. The low level of noise found in this study
for Corbino disc geometry implied that the noise in the samples
of bar geometry was produced essentially at their edges. This
finding can be understood in terms of the surface barriers (of
Bean-Livingston\cite{bl} or geometrical\cite{geom} type)  for a
vortex entering and leaving a sample. In short, a depinned vortex
bundle "shoots" accross a bar-geometry sample, interacts only
weakly with the pinning centres and the rest of (pinned or slowly
moving) vortices, and the noise is created by the bundle
overcoming the barriers at the entry and exit from the sample. In
this case $\Gamma$ does not depend on sample width. However, in
later studies\cite{habjoin,knoedler} it was found that if vortex
bundles travel a distance $x \ll W$ before their motion is
interrupted by the pinning centres, $\Gamma$ should be inversely
proportional to $W$. The reason for this can be inferred from the
Josephson relation $V = \phi_0   (d \varphi / dt) / 2 \pi$,
where $\varphi$ is the phase of the superconducting order
parameter. A moving vortex causes the phase shift of $2 \pi$ only
if it moves over the whole distance $W$. If the actual distance
$x$ is shorter than $W$, the phase change associated with one
voltage pulse is a factor $x/W$ less than $2 \pi$, and the
consequence is $\Gamma = 2  \phi_0 N x / W$.\cite{knoedler} Note
that in this case the noise is produced in the bulk, i.e. at the
pinning centres. The reduction factor $x/W$ explains the result
of Ref.\onlinecite{gurp} that the noise produced in the bulk (by
slow vortices moving over small distances,  or by local
bundle-velocity fluctuations) was much smaller than that due to
the "shooting" bundles overcoming the surface barriers. A more
complicated expression for $\Gamma$ was derived in
Ref.\onlinecite{habjoin}, where it was found that if there is a
distribution of the strengths and positions of pinning centres
the above expression becomes $\Gamma = 2  \phi_0 \langle N^2
\rangle \langle x^2 \rangle / \langle N \rangle \langle x \rangle
W$, where the brackets denote averages over the distribution
function.

In the upper inset to Fig.2 we plot $\Gamma(B)$ 
for sample S5 and $\Gamma (B) / 8$ for sample S1.
Over the whole field range where we found well
defined linear $\Sigma_V (V)$ curves the slopes $\Gamma(B)$ for both samples
decrease with increasing field in the same manner, and $\Gamma$
for sample S1 is approximately eight times larger. If we take
into account slightly different experimental conditions for the
two samples this is in fair agreement with $\Gamma \propto 1/W$.
At magnetic fields lower than $\sim 0.20$ T the resistances of
the samples, the measured voltage and the corresponding voltage
noise  are small, which leads to a larger error in $\Gamma$.

We now address the question of whether the noise is produced by the
pinning or by the surface barriers. The surface barriers are
important at applied magnetic fields of the order of, or lower than,
the thermodynamic critical field $H_c = B_{c2} / \mu_0 \kappa
\sqrt{2}$. In our case, $\mu_0 H_c \sim 11$ mT is much lower than
the fields at which we found the noise of a measurable magnitude.
In addition, the approximate scaling of $\Gamma$ with
sample width suggests that the bulk pinning, and not the sample edges, 
dominates the noise. 
In turn, measurements of the width
dependence of $\Gamma$ may be an alternative to other
experiments\cite{richard} for determining whether the surface
barriers influence the measured transport properties.

The fact that our measurements allow us to exclude the surface
barriers as the main origin of the noise in our samples also
sheds more light on the nature of the $B_{irr}$ and the meaning
of the potential $U$ of HVM. It is known that for some samples
(e.g. single crystals of the Bi$_2$Sr$_2$CaCu$_2$O$_{8+x}$
high-$T_c$ superconductor) surface barriers may have considerable
effect on both the irreversibility field\cite{zeldov} and the
thermally activated transport.\cite{richard} This is not the case
in the present situation, the $B_{irr}$ can be attributed to a
transition to a depinned vortex state and the $U$ is related to
bulk pinning, as we have anticipated in Section \ref{average}.

We attribute the decrease of $\Gamma$ with
increasing $B$ to the weakening of pinning as $B$ approaches
$B_{irr}$, since for $B > B_{irr}$ we found no linear $\Sigma_V
(V)$ curves and, moreover,  the overall noise magnitude decreases
as $B$ increases towards $B_{irr}$. The decrease of $\Gamma
(B)$ {\it for} $B$ {\it well below} $B_{irr}$ 
could be explained within the framework of
the models of Refs.\onlinecite{habjoin,knoedler}, if the unknown
parameters $x$, $N$, and, respectively, $\langle x \rangle, \langle
x^2 \rangle, \langle N \rangle$ and $\langle N^2 \rangle$ depend
on magnetic field in the right way. 
Since $\Gamma$ comprises these parameters as products and ratios 
(see above), they cannot be extracted independently from our data.
However, both models break down in the limit $B \rightarrow
B_{irr}$. This can be understood as follows. The effects of pinning are
(1) formation of vortex bundles in order to increase the driving force and 
thus facilitate their motion against the pinning potential, 
(2) reduction of the hopping distance below the sample width $W$.
The pinning force vanishes at $B_{irr}$, implying 
$\langle N^2 \rangle ^{1/2}
\rightarrow \langle N \rangle \rightarrow 1$ and $\langle x^2
\rangle ^{1/2} \rightarrow \langle x \rangle \rightarrow W$, 
i.e. $\Gamma \rightarrow 2 \phi_0$, which is in contrast to
the experimental observation shown in Fig.2.

A reason for this breakdown of the classical models can possibly
be inferred from the comparison of the transports of (normal) electrons and vortices 
close to the limit of  perfect transmission.
Our experimental realisation - where vortices are created at the
entry into a sample and vanish at the exit, is equivalent to a
two-terminal mesoscopic conductor - where electrons have their
source and drain in the reservoirs. 
Whenever the transmission coefficient $\Theta$
for electron transport through such a mesoscopic conductor is close to unity,
shot noise is suppressed by a factor $(1 - \Theta)$.\cite{buetles} 
In the ballistic limit ($\Theta = 1$) there is no noise associated with
electron transport.  
If vortices are not slowed down by bulk pinning and/or surface
barriers, their motion is determined by the viscous drag only.
This situation represents perfect vortex motion, conceptually 
similar to ballistic transport of electrons. Therefore, if there 
are no dynamic effects present (see Section \ref{highnoise}),
in the limit of perfect vortex transmission accross a sample
the noise should vanish.
A more quantitative
treatment of vortex motion shot noise at high transmittance 
requires further research.

\subsection{\label{highnoise} Noise in the regime of linear $V(I)$}

Above $B_{irr} \sim 0.65$ T, where  the vortex density is large and
$V \approx R_{FF} I$ for all our noise measurements,  no noise described in Section \ref{lownoise}
was found.
Instead, as we show in Fig.3a, $\Sigma_V$
is a monotonic function of $V$, increasing as $V^2$, and
as a function of magnetic field it decreases as $B$ approaches $B_{c2}$.
Moreover, as shown in Fig.3b, there is a scaling
$\Sigma_V  \propto (1 - B/B_{c2})^2 V^2$
which holds over $B_{irr} < B < B_{c2}$
and is insensitive to variations of  $B_{c2}$ in the
range $1.14 - 1.22$ T.
The frequency dependence of $\Sigma_V$ in this regime
is of $1/f$ type, more precisely $1/f^\alpha$ with $\alpha = 1.5 \pm 0.1$
(Fig.3b, upper inset). In the normal state, above $B_{c2}$,
$\Sigma_V = 0$ and $S_V$ is simply the voltage-independent Nyquist noise.

%
%
\begin{figure}
\includegraphics[width=75mm]{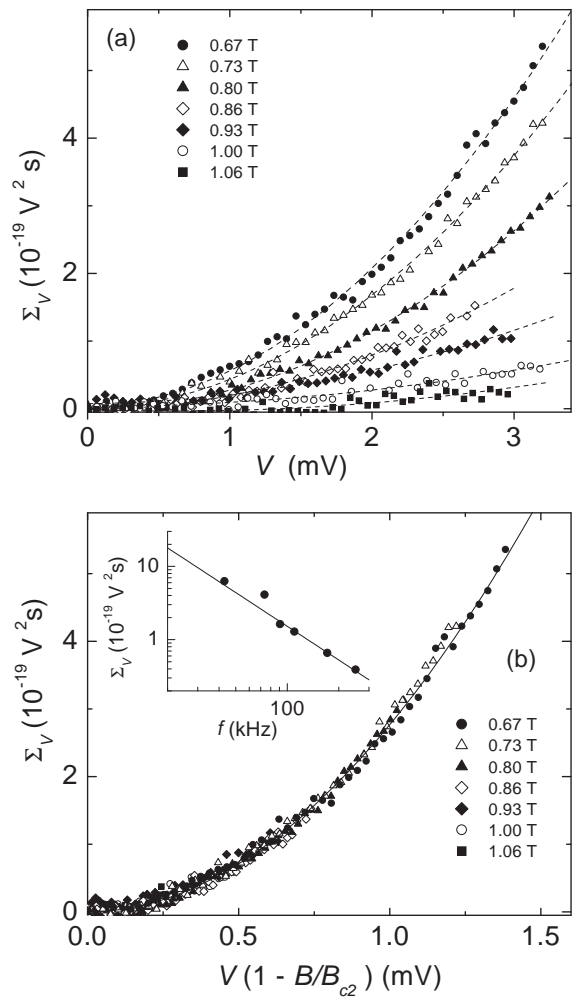}
\caption{(a) Vortex motion noise $\Sigma_V (V)$ for 0.67 T $\leq B \leq$ 1.06 T.
The dashed lines are fits to $\Sigma_V \propto V^2$ dependence.
(b) The curves from (a) plotted against $(1 - B/B_{c2}) V$. Solid line:
$\Sigma_V = \gamma (1 - B/B_{c2})^2 V^2$  with $\gamma (110 \; {\rm kHz}) = 2.1 \times 10^{-13}{\rm s}$.
Upper inset: Frequency dependence of this noise, measured at 0.67 T and 1.5 mV, showing
$\Sigma_V(f) \propto f^{- \alpha}$ with $\alpha = 1.5 \pm 0.1$, as indicated by the solid line.}
\end{figure}

The existence of any noise in the regime where the vortices
are most likely to be completely depinned, as seen from the $R(B,T)$ and
$V(I)$, is rather surprising, since in the pinned state the magnitude
of the noise described in Section \ref{lownoise}
is becoming progressively smaller as $B \rightarrow B_{irr}$.
Furthermore, if the background pinning
would still influence the noise significantly
one would not expect  an increase of $\Sigma_V$ with increasing $V$,
because at larger driving force the role of pinning is less
important. Therefore, the origin of the noise shown in Fig.3
has to be sought in dynamic properties of
depinned vortices far from equilibrium, with a guideline along the LO theory of
non-equilibrium phenomena in flux-flow dissipation.\cite{lo}
In addition, a possible
partial or complete melting of the vortex lattice,
which could occur at $B_{irr}$,\cite{dbjrc}
should also be taken into account.

There is experimental evidence in support of our
assumption that the peculiar noise observed is
not related to depinning processes.
In Fig.4 we show $\Sigma_V(V)$ for
0.61 T, i.e. just at the crossover from HVM to
LO FF in $R(B,T)$. For low voltages, $\Sigma_V (V) \propto V$ (as indicated by the
dashed line), suggesting
that the vortices undergo the HVM.
At $V \sim 0.8$ mV the noise starts to deviate from the linear dependence,
showing in a small voltage range a tendency to decrease,
typically for the vortex motion becoming more ordered with increasing
driving force. However, at higher $V$ the decrease of $\Sigma_V (V)$ does not
continue but instead $\Sigma_V$ approaches the same $\Sigma_V (V) \propto V^2$ behaviour
as for the higher fields (the solid line in Fig.4 indicates the scaling in Fig.3b).
Although the vortex motion is becoming more and more
uniform the noise increases, which can hardly be explained
in terms of vortex interaction with a
pinning potential.

Quadratic voltage dependence and $1/f$ power spectrum are generally known
to be the properties of resistance fluctuations.\cite{weissman}
Hence, a possibility that our finding represents
resistance fluctuations, i.e. vortex velocity fluctuations, requires attention.
At a fixed $(B,T,I)$ point, two parameters influence
vortex velocity and consequently FF resistance:
$\rho_N$ and vortex core area
$A_c$. Thus, if there are fluctuations in either $\rho_N$ or $A_c$, the FF
resistance fluctuates as well.
The fact that the measured noise above $B_{c2}$ is just
the Nyquist noise rules out fluctuations of $\rho_N$, leaving
us with a possibility that
$A_c$ fluctuates.
We argue below that such fluctuations may occur if the vortex velocity is large
and the vortex density high.

%
%
\begin{figure}
\includegraphics[width=65mm]{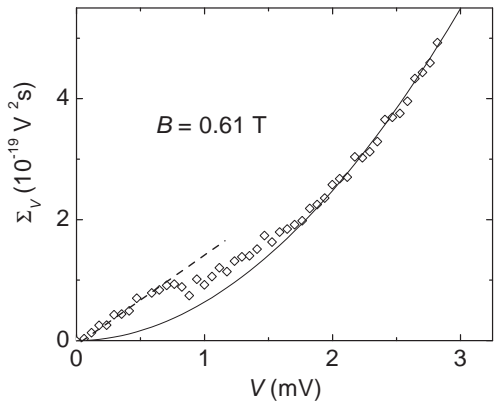}
\caption{Vortex motion noise $\Sigma_V (V)$ for 0.61 T, i.e. for $B$ slightly below $B_{irr}$. For small $V$,
$\Sigma_V \propto V$, as indicated by the dashed line. Above $V \sim 0.8$ mV the noise
in a rather narrow voltage range decreases with increasing $V$, but then increases
again at higher voltages. Eventually, the increase becomes
quadratic in $V$ and approaches the same scaling as for $B > B_{irr}$, shown by the
solid line.}
\end{figure}

The non-equilibrium properties of vortex cores 
and the related influence on flux-flow dissipation
were studied theoretically by LO,\cite{lo} and in
Ref. \onlinecite{bez}.
If the electric field generated in moving vortex cores is
sufficiently strong, quasiparticles in the cores can gain
enough energy to overcome the potential barriers at vortex edges and to escape into
the surrounding superfluid.
This leads to a reduction of the core size,
the vortex viscosity decreases\cite{comvisc} and the vortex velocity increases,
resulting in the non-linearities in $V(I)$ at large currents and finally the jump shown in the
upper inset to Fig.1.
At low vortex density
the electron-phonon relaxation processes are sufficiently
efficient to cool the hot quasiparticles to the bath temperature, as the heating occurs in the cores only
and the cooling over the whole volume.

However, the situation changes at {\it large vortex density}.
With increasing vortex density the
cooling efficiency decreases and the quasiparticles
are heated-up to an elevated temperature.\cite{lo,bez}
This may cause an increase of thermal fluctuations of the
quasiparticle density. As a consequence, the quasiparticle pressure on the vortex "walls"
may fluctuate, which would then result in the fluctuations of $A_c$. The related
fluctuations of $v_\phi$ are measured as voltage fluctuations.

Since the average transport properties can be for $B > B_{irr}$
consistently described by the LO theory, it
is tempting to check whether the LO expression for $\sigma_{FF}$
(see Section \ref{average}) allows to relate the possible
core-size fluctuations and the measured fluctuations in voltage.
Because the observed noise occurs where $V(I)$ is linear, the
$\Sigma_V \propto V^2$ dependence can be explained by assuming
fluctuations of the {\it conductivity}, i.e. $\Sigma_V\Delta
f=(\delta V)^2 = (\delta \sigma_{FF})^2 / \sigma_{FF}^2\;V^2$.
$\Delta f$ is the frequency interval over which the noise
spectrum is averaged. To relate the fluctuations $\delta
\sigma_{FF}$ and $\delta A_c$ we can rewrite $\sigma_{FF}$ in
terms of the vortex core area $A_c \sim \xi^2 \sim \phi_0 /
B_{c2}$ and the intervortex distance $l_B \sim \sqrt{\phi_0 /B}$, so
that $z = B/B_{c2} = A_c / l_B^2$. Then we calculate $\delta
\sigma_{FF} = (1/l_B^2) (\partial \sigma_{FF} /
\partial z) \delta A_c$ from Eq.~\ref{LOFF} and
obtain 
\begin{equation}
\frac{(\delta \sigma_{FF})^2}{\sigma_{FF}^2} = G(B/B_{c2}, T/T_c)\;
\frac{(\delta A_c)^2}{A_c^2}\;\;,
\label{LOFF_noise}
\end{equation}
where $G(z, T/T_c) =[dg(z)/dz - g(z)/z]^2/ [(1- T/T_c)^{1/2} +
g(z)/z]^2$.

The form of $(\delta A_c)^2/A_c^2$ is not known {\it a priori}.
However, it can be deduced by combining Eq.\ref{LOFF_noise}
and the experimentally observed behaviour
$(\delta \sigma_{FF})^2 / \sigma_{FF}^2 = (\delta V)^2/V^2 = \gamma (1 - B/B_{c2})^2  \Delta f$ 
(see Fig.3b). This results in
\begin{equation}
\frac{(\delta A_c)^2}{A_c^2} =
\gamma \frac{(1 - B/B_{c2})^2\;\Delta f}{G(B/B_{c2},
T/T_c)} \;\; .
\label{deltaAc}
\end{equation} 

In Fig.5 we plot this expression against $B/B_{c2}$ in order to
check whether there is any approximation that would lead to a
simple picture of the fluctuations. 
It is seen that  $(\delta A_c)^2/A_c^2$ can be well approximated 
for $B/B_{c2} \lesssim 0.92$ by a power law, 
i.e., $(\delta A_c)^2/A_c^2\propto (B/B_{c2})^{-n}$ 
with $n \approx 2$. 
The simulations for other values of $T/T_c$ show that
the power-law approximation holds well for
essentially any value of $T/T_c$. The power $n$ weakly depends on
$T/T_c$ but is reasonably close to 2 
in the region $0.7 < T/T_c <0.95$. 

The apparent $(B/B_{c2})^{-2}$-decrease of $(\delta A_c)^2/A_c^2$
has a simple visualisation: such a functional dependence 
corresponds to a plausible assumption that the fluctuations
$\delta A_c$ of the vortex area are proportional to the space
$\sim l_B^2$ available, so that $(\delta A_c)/A_c \propto
(l_B^2/\xi^2)\propto(B/B_{c2})^{-2}$.  The above modelling 
based on the LO conductivity hence shows that the 
assumption of core-size fluctuations may reproduce
the measured voltage and magnetic field
dependences of the voltage noise. 

%
%
\begin{figure}
\includegraphics[width=65mm]{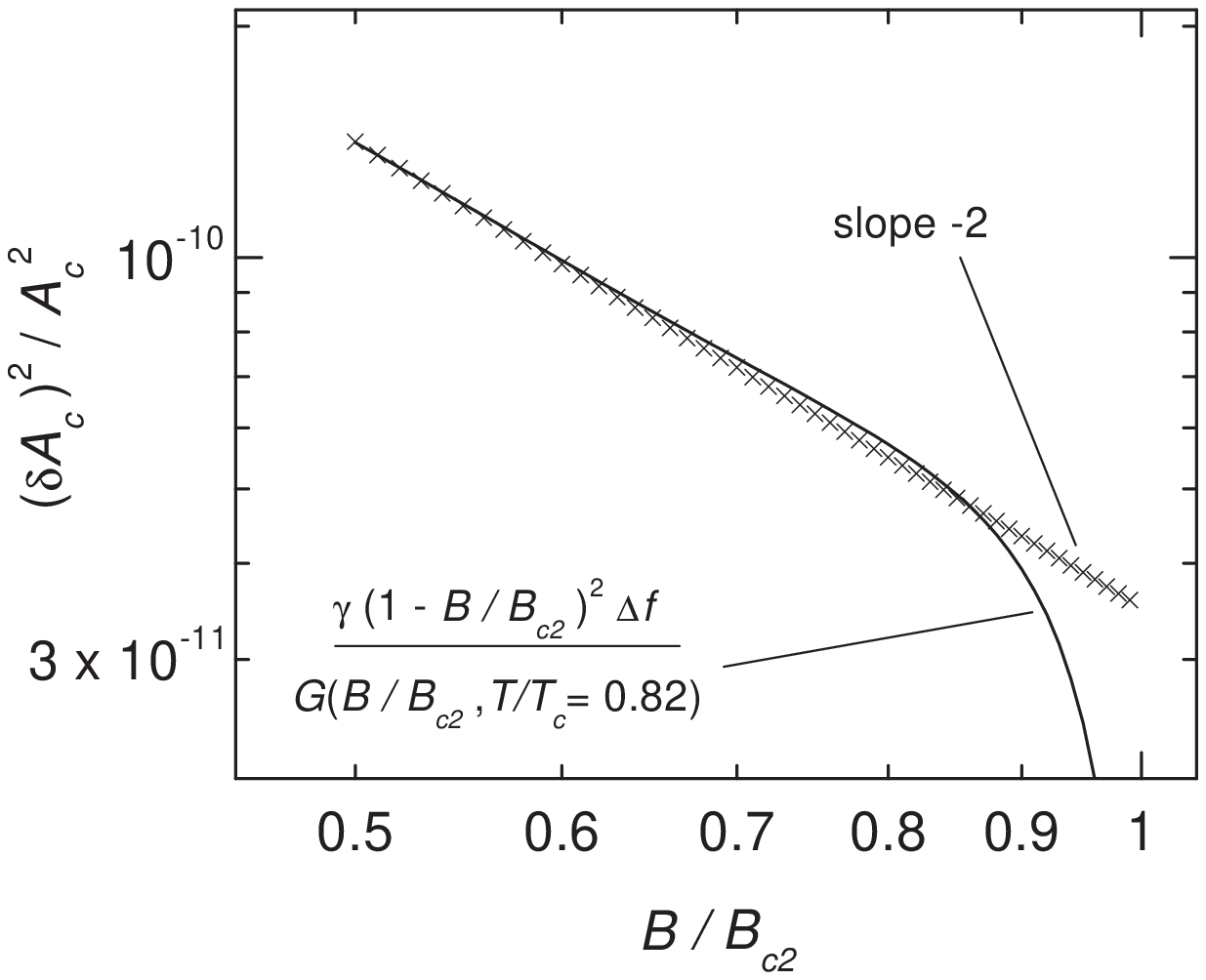}
\caption{Solid line: log-log plot of $(\delta A_c)^2/A_c^2$ 
given by Eq.\ref{deltaAc}, for
$f=110$ kHz, $\Delta f = 7.5$ kHz, $T/T_c = 0.82$,
$\gamma = 2.1 \times 10^{-13}$ s (corresponding
to the data shown in Fig.3b). 
Crosses indicate the $(B/B_{c2})^{-2}$ approximation of
 $(\delta A_c)^2/A_c^2$, discussed in text.}
\end{figure}

With the experimentally determined
value of the prefactor $\gamma = 2.1 \times 10^{-13}$ s for $f = 110$ kHz and
$\Delta f = 7.5$ kHz we obtain the relative fluctuation
amplitude $\delta A_c / A_c$ of the order of $10^{-5}$. 
However, as we discuss below, the $1/f$ spectrum
implies that the fluctuations are distributed over a range
of relaxation times. As a consequence, the small
value of  $\delta A_c / A_c$ only represents the
contribution of those core-size fluctuations which
occur in this frequency window
around the given frequency. 

The observed $1/f$ spectrum cannot be explained if all
vortex cores fluctuate in exactly the same manner. 
The fluctuation of the size of a vortex core is assumed to be a random
process with a characteristic time $\tau$. If $\tau$ would be the same
for all cores, this would result in a Debye-Lorentzian spectrum of
the fluctuations, white up to the cutoff frequency $1 / \tau$.
On the other hand, a distribution of $\tau$ and a superposition
of Debye-Lorentzian spectra
may result in a $1/f$ spectrum.\cite{weissman}
Properties of the distribution then also determine how much the  
fluctuations with a given $\tau$ contribute to $\delta A_c / A_c$ 
measured at $(f,\Delta f)$.
Such a distribution may arise, for example, as a consequence
of {\it different local correlations}.  

That vortex motion can strongly depend on local
conditions was demonstrated in Ref.\onlinecite{matsuda}, where
it was found that, in the presence of pinning, vortices move in a form
of intermittent "rivers" between the pinned islands.
In our case one can hardly discuss a motion around the pinned islands,
since any important pinned fraction would affect the average transport properties
significantly, which is not observed (see the discussion of Fig.1).
This however
does not necessarily imply that there are no "floating islands", i.e.
vortex lattice domains moving together with the "liquid" phase.
The average flux-flow dissipation in such a (depinned) system would
be still well described by the LO theory, since the ratio
$B/B_{c2}$ influences the magnetoresistance much more
strongly than the exact geometry of a system of
moving vortices.\cite{lo}
However, {\it  the local vortex correlations could be different}
for vortices deeply in the
islands, in the "liquid", close to the island boundaries, etc.,
which could lead to different relaxation times for the core fluctuations.
These different relaxation times would then give a $1/f$ noise spectrum.

We are aware that our arguments offer only a
qualitative picture, and that further clarification of the above
ideas is required. However, at the moment we do not know of any
quantitative theoretical model which would account for the
observed peculiarities of vortex motion noise above $B_{irr}$,
nor are we aware of any related systematic experimental work
dealing with a range $B_{irr} < B < B_{c2}$ as large as $\sim 50$
\% of $B_{c2}$. Thus, we believe that the results and discussion
of this Section could be used as a possible starting point for
further experimental and theoretical work.

\section{\label{summary} Summary and conclusions}

We have measured voltage noise in the mixed state of
micrometre-sized wires of amorphous Nb$_{0.7}$Ge$_{0.3}$ thin films.
The samples are well described by conventional
theories for dirty weak-coupling superconductors, have weak
pinning, relatively low irreversibility field $B_{irr}$,
and the vortex structure is much simpler than in high-$T_c$
superconductors. These properties make the samples
suitable for exploring the vortex motion noise in the weak-pinning regime.

At low magnetic fields, i.e. for $B < B_{irr}$, and small applied
currents the voltage-current curves exhibit properties
characteristic of thermally activated hopping of vortices. The
related noise is a linear function of voltage, with the slopes
$\Gamma$ of noise vs. voltage curves  inversely proportional to
the sample width, and is basically frequency independent up to
250 kHz. This behaviour is in agreement with the shot noise model
and the assumption that the noise is generated by bulk
pinning and not by surface barriers. $\Gamma$  decreases with
increasing $B$ over the whole magnetic field range 
of the shot-noise-like behaviour, which does not contradict the presently 
available models of vortex motion shot noise. These models
however fail to explain the disappearance of the shot noise as
$B \rightarrow B_{irr}$.  For $B < B_{irr}$  but at
larger currents the vortex motion becomes more uniform and the
noise decreases. The decrease and the low level of the noise is
ascribed to the ordering of vortex motion with increasing driving
force.

In a narrow range of $B$ slightly below $B_{irr}$, at low $V$ one
still observes the above-mentioned two types of noise but at
large $V$ the noise becomes quadratic in $V$. This signifies the
appearance of the dynamic effects inherent to large vortex
density, a behaviour fully developed for $B > B_{irr}$. For $B > B_{irr}$ the $V(I)$
curves are linear over the whole range of our measurements and
the magnetoresistance agrees well with the flux-flow theory of
Larkin and Ovchinnikov. The noise in this regime is completely
different from that for $B < B_{irr}$. Over the whole voltage
range it increases quadratically with increasing voltage, its
frequency spectrum is of $1/f$ type, and it scales with
$(1-B/B_{c2})^2 V^2$. The origin of this noise is not entirely
clear. We present a qualitative explanation in terms of the
non-equilibrium properties of moving vortex cores which are
subjected to fluctuations of their radius. 

We are greateful to J.~Aarts for contributing to this work in its
initial stage. Valuable discussions with K.~E.~Nagaev,
V.~B.~Geshkenbein, G.~B.~Lesovik, G.~Blatter,
H.~v.~L\"{o}hneysen, F.~Nori, E.~H.~Brandt and J.~R.~Cooper are
greatfully acknowledged. This work was supported by the Swiss
National Science Foundation.


\begin{thebibliography}{99}
%
\bibitem{gurp} D. J. van Ooijen, and G. J. van Gurp, Phys. Lett. {\bf 17}, 239 (1965);
        G. J. van Gurp, Phys. Rev. {\bf 166}, 436 (1968).
\bibitem{clem} For an overview see: J. R. Clem, Phys. Rep. {\bf 75}, 1 (1981). This paper
              also contains details of theoretical approach to vortex motion noise.
\bibitem{woltgens} P. J. M. W\"{o}ltgens, C. Dekker, S. W. A. Gielkens and H. W. de Wijn,
              Physica C {\bf 247}, 67 (1995).
\bibitem{danna} G. D'Anna, P. L. Gammel, H. Safar, G. B. Alers and D. J. Bishop,
    J. Giapintzakis and D. M. Ginsberg,
                Phys. Rev. Lett. {\bf 75}, 3521 (1995).
\bibitem{kim} D. H. Kim, K. E. Gray, N. Jukam, D. J. Miller, Y. H. Kim, J. M. Lee, J. H. Park and
    T. S. Hahn,  Phys. Rev. B {\bf 60}, 3551 (1999).
\bibitem{bl} C. P. Bean and J. D. Livingston, Phys. Rev. Lett. {\bf 12}, 14 (1964).
\bibitem{geom} E. Zeldov, A. I. Larkin, V. B. Geshkenbein, M. Konczykowski, D. Majer,
    B. Khaykovich, V. M. Vinokur and H. Shtrikman,
    Phys. Rev. Lett. {\bf 73}, 1428 (1994).
\bibitem{richard} D. T. Fuchs, R. A. Doyle, E. Zeldov, S. F. W. R. Rycroft, T. Tamegai,
    S. Ooi, M. L. Rappaport and Y. Myasoedov,
    Phys. Rev. Lett. {\bf 81}, 3944 (1998); S. F. W. R. Rycroft, R. A. Doyle,
    D. T. Fuchs, E. Zeldov, R. J. Drost, P. H. Kes, T. Tameagi, S. Ooi and
    D. T. Foord, Phys. Rev. B {\bf 60}, R757 (1999).
\bibitem{aarts} J. M. E. Geers, C. Attanasio, M. B. S. Hesselberth, J. Aarts and P. H. Kes,
    Phys. Rev. B {\bf 63}, 094511 (2001).
\bibitem{theun} M. H. Theunissen and P. H. Kes, Phys. Rev. B {\bf 55}, 15183 (1997); For a
              comprehensive study  see:  M. H. Theunissen, PhD. Thesis, University of Leiden, 1997.
\bibitem{habjoin} F. Habbal and W. C. H. Joiner, J. Low Temp. Phys. {\bf 28}, 83 (1977);
     J. D. Thompson and W. C. H. Joiner, Phys. Rev. B {\bf 20}, 91 (1979).
\bibitem{kes} P. H. Kes and C. C. Tsuei, Phys. Rev. B {\bf 28}, 5126 (1983).
\bibitem{dbjrc} D. Babi\'{c}, J. R. Cooper, J. W. Hodby and Chen Changkang,
             Phys. Rev. B {\bf 60}, 698 (1999).
\bibitem{henny} M. Henny, S. Oberholzer, C. Strunk and C. Sch\"{o}nenberger,
              Phys. Rev. B {\bf 59}, 2871 (1999).
\bibitem{lo} A. I. Larkin and Yu. N. Ovchinnikov, in {\it Nonequilibrium Superconductivity},
              edited by D. N. Lengenberg and A. I. Larkin (North Holland, Amsterdam, 1986).
\bibitem{lefloch} F. Lefloch, C. Hoffmann and O. Demolliens, Physica C {\bf 319}, 258 (1999).
\bibitem{knoedler} C. M. Knoedler and R. F. Voss, Phys. Rev. B {\bf 26}, 449 (1982).
\bibitem{zeldov} E. Zeldov, D. Majer, M. Konczikowski, A. I. Larkin,
    V. M. Vinokur, V. B. Geshkenbein, N. Chikumoto and H. Shtrikman,
    Europhys. Lett. {\bf 30}, 367 (1995).
\bibitem{buetles} G. B. Lesovik, Pis\'ma Zh. Eksp. Teor. Fiz. {\bf 49}, 515 (1989)
    [JETP Lett. {\bf 49}, 594 (1989)];  M. B\"{u}ttiker, Phys. Rev. Lett. {\bf 65}, 2901 (1990).
\bibitem{weissman} For an overview of the various aspects of $1/f$ noise
    and resistance fluctuations see e.g. M. B. Weissman, Rev. Mod. Phys. {\bf 60}, 537 (1988).
\bibitem{bez} A. I. Bezuglyj and V. A. Shklovskij, Physica C {\bf 202}, 234 (1992).
\bibitem{comvisc} From the LO theory it follows that out of equilibrium the
    vortex viscosity is velocity dependent and given by
    $\eta(v_\phi) = \eta_0 [1 + (v_\phi/ v_\phi^*)^2]^{-1}$, where $\eta_0$ is the
    viscosity for small vortex velocities and  $v_\phi^*$ a characteristic vortex velocity.
    Also, $A_c (v_\phi) = A_c(0) [1 + (v_\phi/ v_\phi^*)^2]^{-1}$, which leads to
    $\eta(v_\phi) = \eta_0 A_c (v_\phi) / A_c(0)$. Therefore, one should not confuse
    the vortex-size dependences of   $\eta_0$ and $\eta(v_\phi)$: while  $\eta_0$ is
    larger for smaller $A_c(0)$,  $\eta(v_\phi)$ is smaller for smaller $A_c(v_\phi)$.
\bibitem{matsuda} T. Matsuda, K. Harada, H. Kasai, O. Kanimura and A. Tonomura,
    Science {\bf 271}, 1393 (1996).
%
\end{thebibliography}
\end{document}